%% file: paper.tex
\documentclass[onecolumn,conference]{IEEEtran}
\usepackage[T1]{fontenc}
\usepackage[latin9]{inputenc}
\usepackage{array}
\usepackage{verbatim}
\usepackage{float}
\usepackage{booktabs}
\usepackage{url}
\usepackage{varwidth}
\usepackage{amstext}
\usepackage{graphicx}
\usepackage[]
 {hyperref}

\makeatletter

\providecommand{\tabularnewline}{\\}

\usepackage{cite}
\usepackage{url}
\IEEEoverridecommandlockouts

\usepackage{subfig}
\usepackage{hyphenat}
\hypersetup{
	hidelinks
}

\makeatletter
\renewcommand{\@seccntformat}[1]{\csname the#1\endcsname\quad}
\makeatother

\makeatletter
\def\@IEEEsectpunct{.\ \,}%
\def\section{\@startsection{section}{1}{\z@}%
  {-1.5ex plus -0.5ex minus -0.2ex}%
  {1ex plus 0.2ex}%
  {\normalfont\raggedright\normalsize\scshape\bfseries}}
\makeatother

\usepackage{titlesec}
\titlespacing*{\section}{0pt}{18pt}{0pt}

\makeatother

\begin{document}

\title{Performance Analysis of OpenVPN on a Consumer Grade Router}
\author{\IEEEauthorblockN{Michael J. Hall}\IEEEauthorblockA{Department
of Computer Science \& Engineering\\
Washington University in St. Louis, MO, USA\\
Email: mhall24@wustl.edu}\thanks{A survey paper written for CSE 567M: Computer Systems Analysis, under
the guidance of \protect\href{https://www.cse.wustl.edu/\string~jain/}{Prof. Raj Jain}.
Originally written on November 24, 2008. The original version was
published at: \protect\href{https://www.cse.wustl.edu/\string~jain/cse567-08/ftp/ovpn/index.html}{https://www.cse.wustl.edu/\string~jain/cse567-08/ftp/ovpn/index.html}.}}

\maketitle
\thispagestyle{empty}

\begin{abstract}
Virtual Private Networks (VPNs) offer an alternative solution using
Internet Protocol (IP) tunnels to create secure, encrypted communication
between geographically distant networks using a common shared medium
such as the Internet. They use tunneling to establish end-to-end connectivity.
OpenVPN is a cross-platform, secure, highly configurable VPN solution.
Security in OpenVPN is handled by the OpenSSL cryptographic library
which provides strong security over a Secure Socket Layer (SSL) using
standard algorithms such as Advanced Encryption Standard (AES), Blowfish,
or Triple DES (3DES). The Linksys WRT54GL router is a consumer-grade
router made by Linksys, a division of Cisco Systems, capable of running
under Linux. The Linux-based DD-WRT open-source router firmware can
run OpenVPN on the Linksys WRT54GL router. For this case study, the
performance of OpenVPN is measured and analyzed using a $2^{k-p}$
fractional factorial design for 5 minus 1 factors where $k=5$ and
$p=1$. The results show that the throughput is mainly limited by
the encryption cipher used, and that the round-trip time (RTT) is
mostly dependent on the transport protocol selected.
\end{abstract}


\IEEEpeerreviewmaketitle

\tableofcontents{}

\newpage{}

\input{body.tex}

\bibliographystyle{IEEEtran}
\bibliography{IEEEabrv,refs}

\clearpage{}

\phantomsection
\addcontentsline{toc}{section}{List of Acronyms}

\section*{List of Acronyms}

\begin{tabular}{ll}
\toprule 
\textbf{Acronym} & \textbf{Definition}\tabularnewline
\midrule
3DES & Triple DES\tabularnewline
AES & Advanced Encryption Standard\tabularnewline
CBC & Cipher Block Chaining\tabularnewline
CPU & Central Processing Unit\tabularnewline
DES & Data Encryption Standard\tabularnewline
DNS & Domain Name System\tabularnewline
GPL & General Public License\tabularnewline
ICMP & Internet Control Message Protocol\tabularnewline
IP & Internet Protocol\tabularnewline
LZO & Lempel-Ziv-Oberhumer\tabularnewline
MAC & Medium Access Control\tabularnewline
MMC & MultiMediaCard\tabularnewline
MPEG & Moving Picture Experts Group\tabularnewline
NIC & Network Interface Controller\tabularnewline
NIST & National Institute of Standards and Technology\tabularnewline
OSI & Open Systems Interconnection\tabularnewline
RAM & Random Access Memory\tabularnewline
RFC & Request for Comments\tabularnewline
RTT & Round-Trip Time\tabularnewline
SD & SecureDigital\tabularnewline
SSL & Secure Socket Layer\tabularnewline
SUT & System Under Test\tabularnewline
TAP & Network TAP\tabularnewline
TCP & Transport Control Protocol\tabularnewline
TLS & Transport Layer Security\tabularnewline
TUN & Network TUNnel\tabularnewline
UDP & User Datagram Protocol\tabularnewline
VoIP & Voice over Internet Protocol\tabularnewline
VPN & Virtual Private Network\tabularnewline
\bottomrule
\end{tabular}
\end{document}

%% file: body.tex
\section{Introduction}

In the past, enterprises have used leased lines over long distances
for secure communication between two networks. Typically this is done
in order to communicate data, voice, or other traffic between two
geographically-separated sites of a company or with a valued business
partner. Leased lines provide dedicated bandwidth and a private link
between the two locations. Running leased lines are not always possible
or practical for all enterprises and everyday users due to cost, space,
and time of installation \cite{joha08}. Thus, an alternative solution
is~needed.

Virtual Private Networks (VPNs) were created to address this problem
by using the Internet to facilitate communications. Internet access
is cheap; however, it is insecure and often bandwidth limited. VPNs
are designed to create secure, encrypted Internet Protocol (IP) tunnels
to communicate between geographically-distant networks across the
Internet. This solution is cost-effective for and available to companies
and individuals alike and provides secure access to resources on the
remote~network.

For this case study, the performance of OpenVPN running under Linux
on the Linksys WRT54GL router is analyzed. The sections that follow
give background information on VPNs, and describe the VPN solution
and router used in this case~study. 

\subsection{Background}

\begin{figure}[H]
\begin{centering}
\includegraphics[width=3.5in]{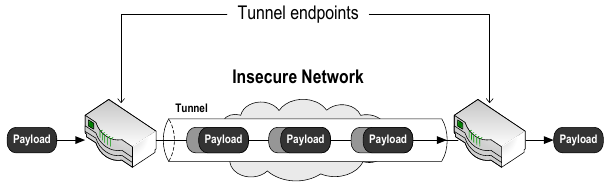}
\par\end{centering}
\caption[VPN tunnel between two endpoints across a network]{\label{fig:tunnel-endpoints}VPN tunnel between two endpoints across
a network~\cite{joha08}}
\end{figure}

Tunneling is a method by which data is transferred across a network
between two endpoints. VPNs use tunnels to establish end-to-end connectivity.
A packet or frame destined to a remote network is first encapsulated
by adding additional header information and is then sent across the
network to the remote endpoint. At the endpoint, the header information
is removed and the packet is sent out onto the remote network \cite{joha08}.
This process is shown in Figure~\ref{fig:tunnel-endpoints}.

There are tradeoffs to using a VPN solution compared to dedicated
lines. A VPN offers benefits such as flexibility, transparency, security,
and cost. However, it has some drawbacks such as availability and
bandwidth \cite{kolesnikov02}. A VPN connection is very flexible
because a user can connect to the remote network from anyplace with
an Internet connection. Transparency is achieved through tunneling
which allows arbitrary traffic to traverse the VPN. For VPNs, security
is provided using authentication and encryption. Authentication restricts
access to the network by allowing only authorized users to connect.
Encryption provides privacy by scrambling the data in the tunnel.
The cost of a VPN is much less than the cost of running dedicated
lines, particularly if a freely available open source VPN solution
is used. VPN solutions are typically deployed to provide access over
the Internet which sometimes varies in the availability and bandwidth
of the connection. In this case, dedicated lines provide a clear advantage.
They are both highly available and provide guaranteed~bandwidth.

In general, a VPN solution should take into consideration security,
key distribution, scalability, transport protocol, interoperability,
and cross-platform availability \cite{kolesnikov02}. Security is
perhaps the biggest concern because there are so many ways to implement
security incorrectly. Key distribution is related to security and
has to do with the procedure by which keys are distributed to clients.
If keys are distributed in an insecure manner, they can be intercepted,
allowing an intruder to gain access to the private network. Scalability
refers to how well a VPN solution scales in terms of the number of
connections and sites. The transport protocol has an effect on the
overhead and performance of the VPN tunnel which will be described
in a later section. Interoperability refers to devices running the
same VPN solution being able to work with each other. Simple, well
thought out designs tend to be the more interoperable. Last, cross-platform
availability allows the VPN solution to work with multiple operating
system~platforms.

In the next section, the VPN solution used in this case study which
takes these points into consideration is~described.

\subsection{The VPN}

OpenVPN is a cross-platform, secure, highly configurable VPN solution
\cite{openvpn}. It uses virtual interfaces provided by the universal
Network TUNnel/TAP (TUN/TAP) driver and is implemented entirely in
user-mode in the least privileged protection ring of the system. This
decision was made to provide better security. If a vulnerability is
found by an intruder, their access will be limited. However, this
does affect performance due to multiple memory copies between kernel
and user space. OpenVPN supports peer-to-peer and multi-client server
configurations which makes many VPN topologies possible: host-host,
host-network, and network-network. It supports creating a Layer 3
or Layer 2 VPN using TUN/TAP devices, respectfully~\cite{feilner06}.

Security in OpenVPN is handled by the OpenSSL cryptographic library
\cite{openssl} which provides strong security over Secure Socket
Layer (SSL) using standard algorithms such as Advanced Encryption
Standard (AES), Blowfish, or Triple DES (3DES). Certificates are used
for authentication, and symmetric and asymmetric ciphers for encryption.
A cipher has several characteristic parameters: key length, block
size, and mode. Key length dictates the strength of the cipher. The
block size dictates how much data is encrypted in a block. The mode
dictates how the encryption cipher is actually used. Other important
factors are key distribution and the cryptographic strength of the
cipher. OpenSSL uses symmetric and asymmetric ciphers as part of the
overall security. However, the security is only as strong as the weakest
link. Kolesnikov and Hatch give an example. If a 40-bit symmetric
key and a 4096 bit asymmetric key are used for the ciphers, likely
the 40-bit key will be the weakest link, making a 4096 bit asymmetric
key unnecessarily large~\cite{kolesnikov02}.

In OpenSSL, block ciphers are used for symmetric encryption and can
be used in different modes. OpenVPN uses a mode called Cipher Block
Chaining (CBC) which makes the cipher text of the current block dependent
on the cipher text of the previous block. This prevents an attacker
from seeing patterns between blocks with identical plaintext messages
and manipulating one or more of these blocks~\cite{kolesnikov02}.

The philosophy for judging the security of an encryption cipher is
based on the test of time. A cipher that has stood scrutiny of the
security community for many years with its details published is generally
considered strong. If the cipher had any major flaws, they likely
would have been found. Some of the criteria for selecting a cipher
are security, performance, and availability. The cipher selected should
meet security needs~\cite{kolesnikov02}.

The cross-platform support in OpenVPN allows it to be deployed to
other systems including embedded routers. The router used in this
case study is described~next. 

\subsection{The Router}

The Linksys WRT54GL router is a consumer-grade router made by Linksys
\cite{linksys}, a division of Cisco Systems, capable of running under
Linux. A Linux firmware actively being developed is DD-WRT \cite{ddwrt}
which is based on the OpenWrt kernel \cite{openwrt}. DD-WRT is released
under the GNU General Public License (GPL) and provides an alternative
to the stock firmware in the Linksys router. The firmware allows the
router to take on many roles: Internet gateway, VPN gateway, firewall,
wireless access point, dynamic Domain Name Service (DNS) client, etc.
It has a friendly web interface and supports many features beyond
the router's original capabilities. It supports OpenVPN through special
firmware and can be extended from the console using packages. Console
access is given by both Secure Shell (SSH) and Telnet. For basic and
advanced configurations, tutorials are available~\cite{ddwrttutorials}.

OpenVPN, which is supported in the DD-WRT firmware \cite{openvpn_ddwrt},
can be used on the router in a variety of different ways. First, key
management can be maintained on the router. This allows certificates
to be generated for users from the console; however, due to a limited
amount of flash memory available, this requires a modification to
the Linksys router to add a SecureDigital/MultiMediaCard (SD/MMC)
memory card. A tutorial on the OpenWrt wiki describes how to do this
\cite{mmctutorial}. Second, the VPN package can be configured in
several different topologies such as host-network or network-network
depending on whether users will connect to the VPN or will access
the remote network using a site-to-site connection. Last, the VPN
virtual interface can be bridged to the physical network interface,
allowing Ethernet frames to traverse between clients and the private~network.

VPNs provide a cost-effective alternative solution to leased lines
and are able to create secure connections between two end-points.
OpenVPN is a VPN solution which can run on an embedded router running
Linux. The Linksys WRT54GL router can run Linux through a firmware
upgrade to the DD-WRT firmware. In the next section, the characteristics
of a VPN will be discussed as they relate to performance~analysis.

\section{VPN Characteristics}

There are many characteristics of a VPN that affects the performance
of the system. In the sections that follow, the transport protocol
is discussed, a set of performance metrics are defined, and the system
parameters are~identified.

\subsection{\label{subsec:Transport-Protocol}Transport Protocol}

The transport protocol used for the VPN tunnel will have an impact
on the performance of the VPN. If the Transport Control Protocol (TCP)
is used, it will have an undesirable effect when TCP is used in the
tunnel as well. This is called TCP stacking. TCP is a connection-oriented
protocol that was not designed to be stacked. It assumes an unreliable
medium and retransmits packets when a timeout occurs. TCP uses an
adaptive timeout which exponentially increases to avoid an effect
known as meltdown. The problem occurs when both TCP protocols timeout.
This is the case when the base connection loses packets. TCP will
queue a retransmission and increase the timeout, trying not to break
the connection. The upper-layer protocol will queue retransmissions
faster than the lower layer due to having a smaller timeout value.
This causes the meltdown effect that TCP was originally trying to
prevent. The User Datagram Protocol (UDP), a datagram carrier having
the same characteristics as IP, should be used as the lower layer
protocol~\cite{titz01}. 

\subsection{Performance Metrics}

Network performance is measured using a set of performance criteria
or metrics. For OpenVPN, the service provided is access to the private
network. The response time, throughput, and utilization are used to
characterize the performance of the VPN. In the case that errors occur,
the probability of errors and time between errors should be measured
\cite{jain91}. A list of selected performance metrics are~below:
\begin{enumerate}
\item Overhead
\item Round-trip time
\item Jitter
\item TCP throughput
\item Router CPU utilization
\item Client CPU utilization
\item Probability of error
\item Time between errors
\item Link utilization
\end{enumerate}
Every VPN packet incurs overhead from the encapsulation process. When
a payload is sent through the VPN tunnel, headers/trailers of various
protocols are added to the payload to form a routable packet. Additional
overhead comes from encryption ciphers used to secure the tunnel.
The effects of overhead can be alleviated by using compression to
reduce the amount of data transmitted~\cite{khanvilkar04}.

The overhead in OpenVPN is a function of the interface, transport
protocol, cryptographic algorithm, and compression. The fixed overhead
added to each packet is 14 bytes from the frame header and 20 bytes
from the IP header. The transport protocol, used to form the VPN tunnel,
contributes 8 (32) bytes from the UDP (TCP) header. The cryptographic
algorithm used to secure the tunnel will contribute to the overhead
depending on the algorithm. Part of the overhead includes the hash
from the medium access control (MAC) algorithm, such as MD5 (128-bits)
or SHA-1 (160-bits), and zero padding for block encryption ciphers.
Compression of uncompressible data adds at most one byte of overhead.
Other minor contributions come from sequence numbers and timestamps
that are included to defeat reply attacks~\cite{khanvilkar04}.

The round-trip time (RTT) is the time it takes for a packet to reach
a remote host and return back and is related to the latency of the
connection. Latency through a VPN tunnel is dependent on the machine
hardware, the link speed, and the encapsulation time. Higher latencies
in OpenVPN are caused by multiple copies between kernel and user space,
and the compute-intensive operations of encryption and compression.
Latency can be improved, generally, by using faster hardware and better
algorithms \cite{khanvilkar04}. Jitter is the variation in the latency
of packets received by a remote host. For applications with streaming
connections, jitter can be alleviated by buffering the stream. However,
this adds delay in the connection which is intolerable for some applications
such as Voice over Internet Protocol (VoIP). Low latency and low jitter
are better for these~metrics.

Throughput is a measure of the amount of payload data that can be
transmitted end-to-end through the VPN tunnel. It does not include
the overhead incurred by protocol headers / trailers, and the VPN
tunnel. Similar to the latency, the throughput is limited by the machine
hardware and encapsulation time; although it can be improved by using
faster hardware and better algorithms. Throughput is a critical performance
metric which will limit the number of users whom the VPN can support.
Thus, higher throughput is~better.

The performance of a VPN solution is often limited by the CPU on one
or both of the endpoints which must encapsulate, encode, transmit,
receive, and decode packets. Monitoring the CPU utilization of each
device allows us to identify the bottleneck in the network communication.
For this metric, utilization in the middle of the range is~better.

The link utilization is the ratio of the physical network interface
throughput to the link speed. In this case, the throughput is the
total throughput of all packets transmitted including overhead. This
metric is not directly useful, however, it can be used indirectly
through a calculation to gauge the efficiency of the packets transmitted
through the VPN tunnel. Higher link utilization is better only when
the throughput is also~higher.

Errors can sometimes occur in network communication causing packets
to be lost, corrupted, duplicated, or out of order. When an error
occurs, it is important to know the probability of it happening again,
and the time between errors. A related metric is packet loss which
gives the percentage of packets that were lost or corrupted. No errors
are ideal, but low error rate is~acceptable. 

\subsection{System Parameters}

The performance in OpenVPN is affected by many parameters ranging
from the hardware to the configuration. A list of these parameters
is below.
\begin{enumerate}
\item Network topology
\item Memory
\item Speed of the router CPU
\item Speed of the network
\item VPN topology
\item Interface
\item Transport Protocol
\item Encryption cipher
\item Encryption key size
\item Compression algorithm
\end{enumerate}
The network topology will affect system performance. For example,
a topology in which the client is located far from the OpenVPN server
will have to contend with network traffic unlike a client that is
directly connected. Hardware factors such as memory, speed of the
router CPU, and speed of the network can all affect system performance.
At least one of these three hardware factors will be a bottleneck
in the system. The VPN topology, such as host-host, host-network,
and network-network, will affect system performance. The choice of
the interface, transport protocol, encryption cipher, key size, and
compression algorithm will all affect system performance. This is
due to additional overhead and compute-intensive~operations.

The transport protocol, if it is TCP, for the VPN tunnel will have
an impact on the network performance. UDP is a protocol which has
the same characteristics as IP which will not suffer from meltdown
caused by retransmissions. There are many metrics that are used to
measure network performance. While not all of them are significant,
each one should nonetheless be included in the performance analysis.
There are many system parameters that affect the performance of the
VPN. Only a subset of these parameters can be changed. For the performance
analysis done in the next section, a set of factors were chosen and
varied in a $2^{k-p}$ fractional factorial~design.

\section{Performance Analysis}

There are three evaluation techniques that can be used for performance
analysis: analytical modeling, simulation, and measurement. For assessing
the performance of OpenVPN on the Linksys WRT54GL router, measurement
was chosen. The performance can be measured and analyzed through a
series of experiments using the experimental design method described
by Jain called $2^{k-p}$ fractional factorial design \cite{jain91}.
This method is intended to determine the effects and percent variation
of the effects for each factor and their interactions. Confidence
intervals can also be calculated to determine the significance of
each effect. In the sections that follow, the measurement tools used
and the experimental setup are described. The results of the study
are then~presented. 

\subsection{Measurement Tools}

Assessing the performance of OpenVPN requires the use of several measurement
tools for generating, measuring, and monitoring network traffic. The
tools used in this case study are \textit{wireshark}, \textit{iperf},
\textit{ping}, and \textit{sar}. Of these four tools, \textit{wireshark},
\textit{iperf}, and \textit{ping} are available for both Windows and
Linux. Although Windows has a \textit{ping} tool, it is limited and
not sufficient for the latency and packet loss tests in this case
study. The last tool, \textit{sar}, is available only on Linux. A
description of each tool is given~below.
\begin{itemize}
\item \textit{Wireshark}, formally \textit{Ethereal}, is a network protocol
analyzer with a rich feature set for capturing and analyzing network
traffic. It has deep inspection and filtering capabilities of hundreds
of protocols, making it a valuable tool for monitoring network traffic
\cite{wireshark}. In this case study, it was used to monitor VPN-encapsulated
packets and normal~packets.
\item \textit{Iperf} is a network testing tool for creating and measuring
TCP and UDP streams. It has options for controlling several network
parameters including maximum segment size (MSS), buffer length, TCP
window size, and TCP no delay (for disabling Nagle's Algorithm). The
TCP test will generate traffic at full speed and measure the bandwidth
between two endpoints. The UDP test will generate traffic at a given
bandwidth and measure the jitter (variation in the latency) and packet
loss between two endpoints~\cite{iperf}.
\item \textit{Ping} is a network testing tool for measuring latency and
packet loss between two endpoints using the Internet Control Message
Protocol (ICMP). A large number of packets can be transmitted using
a flood ping. A flood ping works by transmitting one packet at a time
and waiting for a reply or timeout. If a timeout occurs, the packet
is counted as~lost.
\item \textit{Sar} is a system activity collection and reporting tool found
in the sysstat utilities package \cite{sysstat}. It is able to collect
and report information on CPU and network interface activity over
a period of time. This information can be collected in parallel with
the TCP bandwidth, UDP jitter, and latency~tests.
\end{itemize}

\subsection{Experimental Setup}

The goal of this case study is to evaluate the performance of OpenVPN
on a consumer grade router running the DD-WRT firmware. The router
is a Linksys WRT54GL v1.1 with 16~MB RAM, 4~MB flash memory, and
a 200~MHz~processor.

\begin{figure}[H]
\begin{centering}
\includegraphics[width=3.5in]{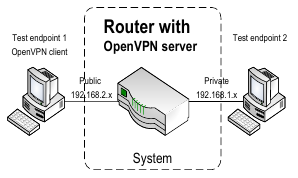}
\par\end{centering}
\caption[System definition for the study of OpenVPN on a consumer grade router]{\label{fig:system-definition}System definition for the study of
OpenVPN on a consumer grade router}
\end{figure}

The system definition, shown in Figure~\ref{fig:system-definition},
consists of two systems connected to a router in the middle. The first
system is the OpenVPN client which needs to establish a VPN tunnel
to access the internal private network. It is the first test endpoint
for the performance tests. The router is the OpenVPN server which
is the system under test (SUT) \cite{jain91}. The second system is
a computer on the private network which is the second test endpoint.
The specifications of these two test systems are shown in Table~\ref{tab:system-specifications}.

\begin{table}[H]
\begin{centering}
\begin{tabular}{V{\linewidth}V{\linewidth}}
\toprule 
\textbf{System} & \textbf{Description}\tabularnewline
\midrule
Test endpoint 1

(test client) & Cento OS 5.2 Linux,

AMD Athlon XP-M 2600+,

1.83 GHz, 2.0 GB of RAM,

ASUS A7V880 motherboard,

VIA KT880 chipset\\\tabularnewline
Test endpoint 2

(test server) & Windows XP Professional, Service Pack 3,

AMD Athlon XP 2000+ (Thoroughbred),

1.67 GHz, 768 MB of RAM,

ASUS A7V8X-X motherboard,

VIA KT400 chipset\tabularnewline
\bottomrule
\end{tabular}
\par\end{centering}
\caption{\label{tab:system-specifications}System specifications for test endpoints}
\end{table}

To facilitate the testing process, a Python script, on test endpoint
1, is used to automatically run each test and collect data results
which are saved in a log file. There are three tests that are run:
TCP bandwidth test, UDP jitter test, and latency test. The TCP bandwidth
test uses \textit{iperf} to generate traffic from a workload, and
to measure the bandwidth over multiple time intervals. Simultaneously,
CPU and network activity are measured using \textit{sar} over the
same time intervals. The UDP jitter test also uses \textit{iperf}
to generate traffic from a workload, and measures the jitter and packet
loss across the VPN tunnel. As with the TCP test, \textit{sar} is
used to measure CPU and network activity. The latency test uses \textit{ping}
to flood small packets to the remote host and to measure the return-trip
time (RTT) and packet loss across the VPN tunnel. The RTT is related
to the latency of the connection, and the packet loss can be used
to get the probability of~error.

Many of the system parameters were fixed to a single value for a reduction
in the number of total experiments needed. Some parameters are determined
by the hardware which cannot be changed such as memory, router CPU
speed, and network speed. The other parameter values were chosen based
on recommendations given in the OpenVPN documentation. These parameters
are shown in Table~\ref{tab:parameters}.

\begin{table}[H]
\begin{centering}
\begin{tabular}{ll}
\toprule 
\textbf{Fixed Parameter} & \textbf{Value}\tabularnewline
\midrule
Memory & 16 MB RAM\tabularnewline
Speed of router CPU & 200 MHz\tabularnewline
Speed of network & 100 Mbps\tabularnewline
Encryption key size & 256-bit\tabularnewline
Digest & SHA1\tabularnewline
TLS cipher & DHE-RSA-AES256-SHA\tabularnewline
\bottomrule
\end{tabular}
\par\end{centering}
\caption{\label{tab:parameters}Parameters fixed in the experimental setup}
\end{table}

In the fractional factorial design, five factors were chosen and are
listed in Table~\ref{tab:factors-ff-design}.

\begin{table}[H]
\begin{centering}
\begin{tabular}{clll}
\toprule 
 & \textbf{Factor} & \textbf{Level (-1)} & \textbf{Level (+1)}\tabularnewline
\midrule
\textbf{A} & Interface & TAP (bridged) & TUN\tabularnewline
\textbf{B} & Protocol & UDP & TCP\tabularnewline
\textbf{C} & Cipher & None & AES-256\tabularnewline
\textbf{D} & Compression & None & LZO\tabularnewline
\textbf{E} & Workload & Text & Video\tabularnewline
\bottomrule
\end{tabular}
\par\end{centering}
\caption{\label{tab:factors-ff-design}Factors in fractional factorial design}
\end{table}

OpenVPN supports creating tunnels using two devices: TUN and TAP.
The primary distinction between these two is the layer at which they
operate. TUN, which stands for Network TUNnel, operates at Layer~3
of the OSI model and will not transmit any Layer~2 protocols through
the VPN. TAP, which stands for Network TAP, operates at Layer~2 of
the OSI model. It is capable of sending Layer 2 protocols through
the VPN, but needs a bridge between the virtual network interface
controller (NIC) and the physical NIC. If bridging mode is not used,
then additional routing table entries are needed to route packets
between the client and remote network~\cite{openvpn_howto}.

Two transport protocols supported by OpenVPN are UDP and TCP. The
performance of the VPN depends on the protocol used. UDP is a datagram
packet which has less overhead and shares the same characteristics
of IP. TCP, however, is a connection-based protocol which assumes
an unreliable medium. Consequently, it has more overhead, and will
encounter adverse effects from packet loss as described in Section~\ref{subsec:Transport-Protocol}.

Many ciphers are available in OpenVPN which uses the OpenSSL cryptographic
library. For this factor, a comparison is being done between no encryption,
and encryption using the Advanced Encryption Standard (AES) algorithm.
This algorithm is recommended by the National Institute of Standards
and Technology (NIST) federal agency for secure communications~\cite{nist}.

In low-throughput links, compression is a way to increase the overall
throughput. The throughput achieved, however, is dependent on the
workload. For this study, the performance is tested both with and
without compression and for different workloads. Two workloads were
chosen: text, and video. The text workload consists of highly compressible
RFC documents and the video workload consists of uncompressible MPEG~video.

The $2^{5-1}$ fractional factorial design for measuring the effects
of each factor is shown in Table~\ref{tab:fractional-factorial-design-2-5-1}.
The design is the equivalent of a 4 factor design, except that the
ABCD interaction is replaced with factor E, the workload. The confounding
effects for this design are shown in Table~\ref{tab:confounding-effects}.

\begin{table}[H]
\begin{centering}
\begin{tabular}{rrrrrrrrrrrrrrrr}
\toprule 
\textbf{I} & \textbf{A} & \textbf{B} & \textbf{C} & \textbf{D} & \textbf{AB} & \textbf{AC} & \textbf{AD} & \textbf{BC} & \textbf{BD} & \textbf{CD} & \textbf{ABC} & \textbf{ABD} & \textbf{ACD} & \textbf{BCD} & \textbf{E}\tabularnewline
\midrule
1 & -1 & -1 & -1 & -1 & 1 & 1 & 1 & 1 & 1 & 1 & -1 & -1 & -1 & -1 & 1\tabularnewline
1 & -1 & -1 & -1 & 1 & 1 & 1 & -1 & 1 & -1 & -1 & -1 & 1 & 1 & 1 & -1\tabularnewline
1 & -1 & -1 & 1 & -1 & 1 & -1 & 1 & -1 & 1 & -1 & 1 & -1 & 1 & 1 & -1\tabularnewline
1 & -1 & -1 & 1 & 1 & 1 & -1 & -1 & -1 & -1 & 1 & 1 & 1 & -1 & -1 & 1\tabularnewline
1 & -1 & 1 & -1 & -1 & -1 & 1 & 1 & -1 & -1 & 1 & 1 & 1 & -1 & 1 & -1\tabularnewline
1 & -1 & 1 & -1 & 1 & -1 & 1 & -1 & -1 & 1 & -1 & 1 & -1 & 1 & -1 & 1\tabularnewline
1 & -1 & 1 & 1 & -1 & -1 & -1 & 1 & 1 & -1 & -1 & -1 & 1 & 1 & -1 & 1\tabularnewline
1 & -1 & 1 & 1 & 1 & -1 & -1 & -1 & 1 & 1 & 1 & -1 & -1 & -1 & 1 & -1\tabularnewline
1 & 1 & -1 & -1 & -1 & -1 & -1 & -1 & 1 & 1 & 1 & 1 & 1 & 1 & -1 & -1\tabularnewline
1 & 1 & -1 & -1 & 1 & -1 & -1 & 1 & 1 & -1 & -1 & 1 & -1 & -1 & 1 & 1\tabularnewline
1 & 1 & -1 & 1 & -1 & -1 & 1 & -1 & -1 & 1 & -1 & -1 & 1 & -1 & 1 & 1\tabularnewline
1 & 1 & -1 & 1 & 1 & -1 & 1 & 1 & -1 & -1 & 1 & -1 & -1 & 1 & -1 & -1\tabularnewline
1 & 1 & 1 & -1 & -1 & 1 & -1 & -1 & -1 & -1 & 1 & -1 & -1 & 1 & 1 & 1\tabularnewline
1 & 1 & 1 & -1 & 1 & 1 & -1 & 1 & -1 & 1 & -1 & -1 & 1 & -1 & -1 & -1\tabularnewline
1 & 1 & 1 & 1 & -1 & 1 & 1 & -1 & 1 & -1 & -1 & 1 & -1 & -1 & -1 & -1\tabularnewline
1 & 1 & 1 & 1 & 1 & 1 & 1 & 1 & 1 & 1 & 1 & 1 & 1 & 1 & 1 & 1\tabularnewline
\bottomrule
\end{tabular}
\par\end{centering}
\caption{\label{tab:fractional-factorial-design-2-5-1}$2^{5-1}$ fractional
factorial design; A = Interface, B = Protocol, C = Cipher, D = Compression,
E = Workload}
\end{table}

\begin{table}[H]
\begin{centering}
\begin{tabular}{cccccccc}
\toprule 
\textbf{I} & \textbf{A} & \textbf{B} & \textbf{C} & \textbf{D} & \textbf{AB} & \textbf{AC} & \textbf{AD}\tabularnewline
ABCDE & BCDE & ACDE & ABDE & ABCE & CDE & BDE & BCE\tabularnewline
\midrule
\midrule 
\textbf{BC} & \textbf{BD} & \textbf{CD} & \textbf{ABC} & \textbf{ABD} & \textbf{ACD} & \textbf{BCD} & \textbf{E}\tabularnewline
ADE & ACE & ABE & DE & CE & BE & AE & ABCD\tabularnewline
\bottomrule
\end{tabular}
\par\end{centering}
\caption{\label{tab:confounding-effects}Confounding effects of $2^{5-1}$
fractional factorial design}
\end{table}

\subsection{Experimental Results}

The results of the $2^{5-1}$ fractional factorial design are presented
in the section below. There are a few general observations that I
made while running these tests. First, the CPU utilization of the
router was always at 100\% during these tests. This indicates that
the router was consistently the bottleneck. The CPU utilization of
the router was not reported because there was no mechanism for obtaining
this information automatically from the router. Second, the CPU utilization
of the client machine was always low, indicating that it was never
the~bottleneck.

The overhead of the VPN tunnel can be estimated empirically using
the network activity information gathered by \textit{sar} during the
TCP test. The overhead is simply the difference in the throughputs
of the physical and virtual network interfaces divided by the number
of packets per second transmitted. The equation is shown as~follows:

\begin{equation}
\text{overhead}=\frac{\text{physical network interface throughput}-\text{virtual network interface throughput}}{\text{packets per second}}\label{eq:overhead}
\end{equation}

For each experiment in the TCP test, traffic was generated from the
client to the server and multiple metrics were measured: bandwidth,
link utilization, and CPU utilization. The results shown in Table~\ref{tab:client-to-server-tcp-test-results}
are the mean value of 5 replications. These results show that the
link was highly underutilized and that the client CPU was not a bottleneck
in these~experiments.

The overhead was estimated only for experiments that did not involve
compression since compression reduces the packet size in the physical
interface. The results show the smallest overhead of 51 bytes for
the UDP protocol without encryption. This number is an estimate of
the overhead due to the encapsulation of packets in the VPN tunnel.
It does not show the overhead due to headers in the payload itself.
For Layer~3 VPNs, which use the TUN interface, the payload does not
contain any Layer 2 header information which reduces the overhead
by 14~bytes.

The largest bandwidth of 8.87~Mbps is measured for the TAP interface
with bridging using the UDP transport protocol for the tunnel and
no encryption. The bandwidth is cut to less than half to 3.70~Mbps
when encryption using AES 256-bit is~enabled.

\begin{table}[H]
\begin{centering}
\begin{tabular}{lllllrrrr}
\toprule 
 &  &  &  &  & \multicolumn{4}{c}{\textbf{TCP Tests}}\tabularnewline
\midrule 
\textbf{Interface} & \textbf{Protocol} & \textbf{Cipher} & \textbf{Comp} & \textbf{Workload} & \textbf{Overhead (Bytes)} & \textbf{BW (Mbps)} & \textbf{Link \%} & \textbf{Client CPU \%}\tabularnewline
\midrule
TAP (br) & UDP & None & None & Video & 51.0 & 8.87 & 1.20\% & 4.12\%\tabularnewline
TAP (br) & UDP & None & LZO & Text &  & 8.02 & 0.70\% & 5.92\%\tabularnewline
TAP (br) & UDP & AES256 & None & Text & 100.0 & 3.64 & 0.51\% & 3.48\%\tabularnewline
TAP (br) & UDP & AES256 & LZO & Video &  & 3.70 & 0.52\% & 3.58\%\tabularnewline
TAP (br) & TCP & None & None & Text & 76.5 & 6.17 & 0.85\% & 3.76\%\tabularnewline
TAP (br) & TCP & None & LZO & Video &  & 6.27 & 0.87\% & 3.90\%\tabularnewline
TAP (br) & TCP & AES256 & None & Video & 127.5 & 3.26 & 0.47\% & 3.54\%\tabularnewline
TAP (br) & TCP & AES256 & LZO & Text &  & 3.83 & 0.36\% & 4.42\%\tabularnewline
TUN & UDP & None & None & Text & 51.0 & 7.38 & 0.99\% & 3.30\%\tabularnewline
TUN & UDP & None & LZO & Video &  & 7.32 & 0.98\% & 3.56\%\tabularnewline
TUN & UDP & AES256 & None & Video & 98.0 & 3.35 & 0.46\% & 3.16\%\tabularnewline
TUN & UDP & AES256 & LZO & Text &  & 3.66 & 0.33\% & 4.16\%\tabularnewline
TUN & TCP & None & None & Video & 76.5 & 5.67 & 0.77\% & 3.28\%\tabularnewline
TUN & TCP & None & LZO & Text &  & 6.11 & 0.53\% & 4.70\%\tabularnewline
TUN & TCP & AES256 & None & Text & 125.0 & 3.05 & 0.43\% & 3.30\%\tabularnewline
TUN & TCP & AES256 & LZO & Video &  & 3.00 & 0.42\% & 3.30\%\tabularnewline
\bottomrule
\end{tabular}
\par\end{centering}
\caption{\label{tab:client-to-server-tcp-test-results}Client-to-server TCP
test results for $2^{5-1}$ fractional factorial design; the values
shown are the mean of 5 replications}
\end{table}

The UDP and latency test results are shown in Table~\ref{tab:udp-and-latency-test-results}.
For both tests, the packet loss percentage is 0\% indicating that
no errors occurred in the VPN tunnel. The jitter measurement was done
using large packets equal to the MTU size with a fixed UDP bandwidth
of 1 Mbps. Although not shown, the jitter measurements were found
to be sensitive to the UDP bandwidth and payload length, but not to
the factors under~test.

\begin{table}[H]
\begin{centering}
\begin{tabular}{lllllrrrrr}
\toprule 
 &  &  &  &  & \multicolumn{2}{c}{\textbf{UDP Tests}} & \multicolumn{3}{c}{\textbf{Latency Tests}}\tabularnewline
\midrule 
\textbf{Interface} & \textbf{Protocol} & \textbf{Cipher} & \textbf{Comp} & \textbf{Workload} & \textbf{Jitter (ms)} & \textbf{Loss \%} & \textbf{RTT (ms)} & \textbf{Loss \%} & \textbf{CPU \%}\tabularnewline
\midrule
TAP (br) & UDP & None & None & Video & 6.2 & 0.00\% & 1.3 & 0.00\% & 7.02\%\tabularnewline
TAP (br) & UDP & None & LZO & Text & 6.2 & 0.00\% & 1.3 & 0.00\% & 6.62\%\tabularnewline
TAP (br) & UDP & AES256 & None & Text & 6.3 & 0.00\% & 2.3 & 0.00\% & 5.22\%\tabularnewline
TAP (br) & UDP & AES256 & LZO & Video & 6.3 & 0.00\% & 2.2 & 0.00\% & 5.34\%\tabularnewline
TAP (br) & TCP & None & None & Text & 6.5 & 0.00\% & 15.1 & 0.00\% & 0.40\%\tabularnewline
TAP (br) & TCP & None & LZO & Video & 6.1 & 0.00\% & 12.7 & 0.00\% & 0.76\%\tabularnewline
TAP (br) & TCP & AES256 & None & Video & 6.3 & 0.00\% & 7.9 & 0.00\% & 2.34\%\tabularnewline
TAP (br) & TCP & AES256 & LZO & Text & 6.6 & 0.00\% & 14.9 & 0.00\% & 0.52\%\tabularnewline
TUN & UDP & None & None & Text & 6.2 & 0.00\% & 1.9 & 0.00\% & 5.14\%\tabularnewline
TUN & UDP & None & LZO & Video & 6.4 & 0.00\% & 1.9 & 0.00\% & 5.04\%\tabularnewline
TUN & UDP & AES256 & None & Video & 6.3 & 0.00\% & 2.8 & 0.00\% & 4.10\%\tabularnewline
TUN & UDP & AES256 & LZO & Text & 6.2 & 0.00\% & 2.8 & 0.00\% & 4.20\%\tabularnewline
TUN & TCP & None & None & Video & 6.2 & 0.00\% & 8.4 & 0.00\% & 1.50\%\tabularnewline
TUN & TCP & None & LZO & Text & 7.0 & 0.00\% & 15.1 & 0.00\% & 0.50\%\tabularnewline
TUN & TCP & AES256 & None & Text & 6.3 & 0.00\% & 11.2 & 0.00\% & 1.84\%\tabularnewline
TUN & TCP & AES256 & LZO & Video & 6.3 & 0.00\% & 10.5 & 0.00\% & 0.82\%\tabularnewline
\bottomrule
\end{tabular}
\par\end{centering}
\caption{\label{tab:udp-and-latency-test-results}UDP and latency test results
for $2^{5-1}$ fractional factorial design; the values shown are the
mean of 5 replications}
\end{table}

\begin{figure}[H]
\begin{centering}
\includegraphics[width=6in]{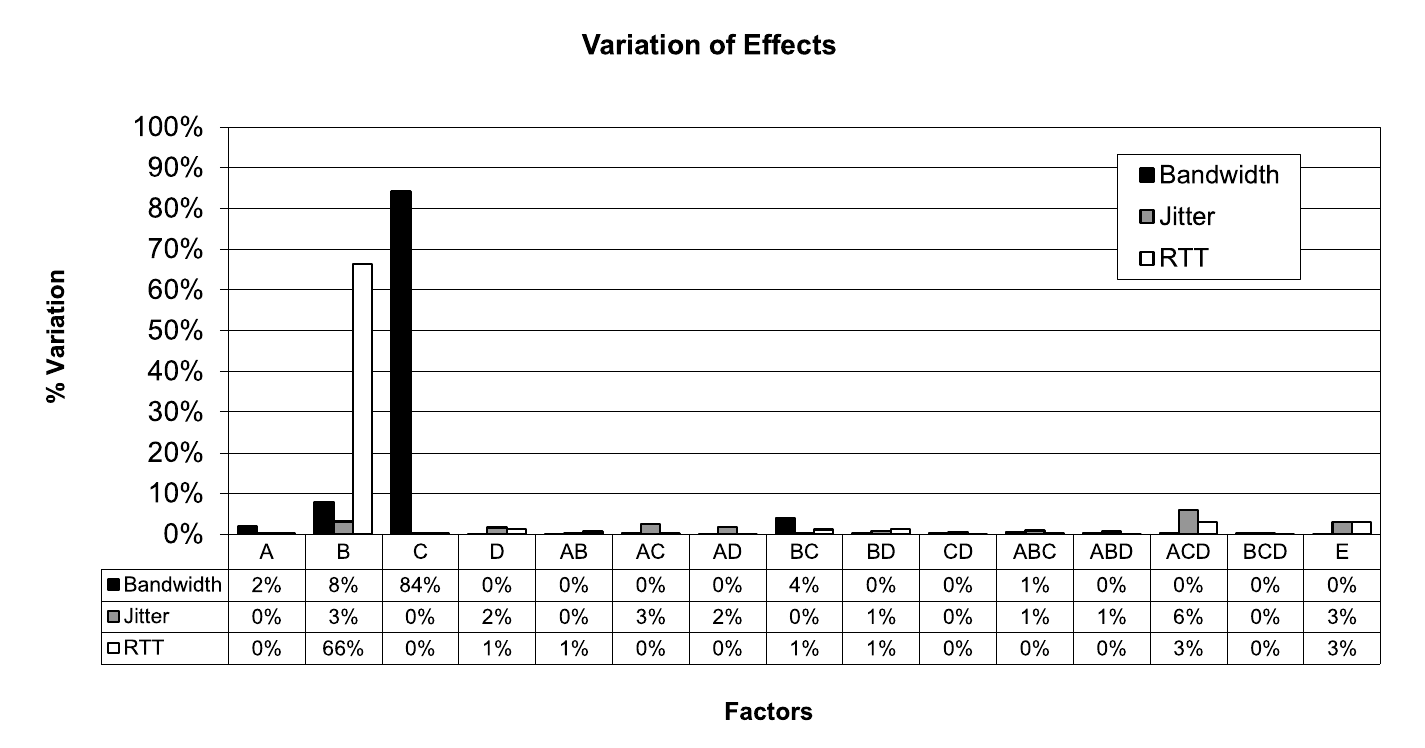}
\par\end{centering}
\caption[Variation of effects for $2^{5-1}$ fractional factorial design]{\label{fig:variation-of-effects}Variation of effects for $2^{5-1}$
fractional factorial design}
\end{figure}

Using the analysis technique described for a $2^{k-p}$ fractional
factorial design, the effects and the percent variation were calculated
\cite{jain91}. The percent variation of the effects is shown in Figure~\ref{fig:variation-of-effects}.
84\% of the bandwidth is explained by encryption cipher (C). Another
8\% is explained by the transport protocol (B). There is a small 4\%
interaction (BC) between the encryption cipher and the transport protocol
of the~bandwidth.

The jitter is not explained very well by the model and is relatively
independent of the changes in the factors. The percent variation in
the round-trip time (RTT) is 66\% for the transport protocol. Another
3\% is explained by the interaction (BE) between the transport protocol
and the workload which is confounded with interaction~ACD.

\subsection{Future Work}

The tests performed in the $2^{k-p}$ fractional factorial design
should be extended to include server-to-client tests, and simultaneous
bidirectional traffic tests. Although it is possible to get this information,
it is not easy to do this in an automatic way, and will require extension
of the test script. The design should also be extended to include
other factors that will potentially affect the performance of the
VPN such as payload length, encryption key size, encryption digest,
TLS digest, etc. Using the results from the $2^{k-p}$ fractional
factorial design, a one or two factor design should be done to analyze
the effects of the most sensitive factors such as the transport protocol
and encryption. The test bed can also be expanded for testing a site-to-site
VPN topology involving two routers. Last, the effects of communication
with multiple clients should also be~tested.

In addition to these tests, the results should be verified using another
evaluation technique such as analytical modeling. Some of the contributions
of the overhead were modeled in this paper, but this work needs to
be expanded to explain changes in performance as well. This will allow
us to understand how each of these factors affects the performance
and~why. 

\section{Summary}

The Linksys WRT54GL router is an inexpensive router that can easily
be upgraded for extended functionality using an open-source Linux
firmware called DD-WRT. This allows a VPN package such as OpenVPN
to be set up to allow remote access to the internal network. OpenVPN
is a flexible, cross-platform solution that is highly configurable
and fairly easy to set up using available~tutorials.

The performance of OpenVPN depends on the router hardware, and the
configuration parameters. The throughput was found to be limited by
the router CPU, and is not sufficient for fast connections such as
10/100 Mbps LANs. It is sufficient for slower connections such as
most Internet connections. Measurements were presented for traffic
generated from client to server. The encryption cipher was found to
significantly reduce total throughput. For a configuration using the
TAP interface with bridging, UDP transport protocol, AES256 cipher,
and no compression, the throughput was 3.64~Mbps. 96\% of the variation
in the throughput was explained by the transport protocol, encryption
cipher, and the interaction between the two; the encryption cipher
explained the majority (84\%) of the variation. The jitter in the
latency was found to be relatively insensitive to the factors tested
at around 6.3~ms. The round-trip time (RTT) was significantly larger
for the TCP transport protocol explaining 66\% of the variation. The
next significant factor was the workload (3\%), followed by the interaction
between the workload and the encryption cipher (3\%). This interaction
is confounded with the interaction between the interface, cipher,
and compression factors. For the same configuration above, the average
RTT was 2.3~ms.

Although the encryption cipher accounted for the majority of the variation
in the throughput, it is an important feature in VPNs. Future work
is to investigate effects of different encryption algorithms with
varying key sizes on the throughput that are still considered strong.
One criterion for choosing an encryption algorithm is whether or not
it is acceptable for use in~ecommerce.

In conclusion, the Linksys WRT54GL router provides a cost-effective
solution for setting up an OpenVPN server for remote access over the
Internet. This solution is throughput-limited, but should be sufficient
for most Internet connections. It is an appropriate solution for most
home users and small businesses depending on their~needs. 

\phantomsection
\addcontentsline{toc}{section}{Data Availability Statement}

\section*{Data Availability Statement}

The data supporting the findings of this study are openly available
and can be accessed from the following GitHub repository: \url{https://github.com/mhall24/openvpn_casestudy_data}.